\documentclass{PoS}
\usepackage{amsmath}

\newcommand{\bra}{\langle}
\newcommand{\ket}{\rangle}

\newcommand{\One}{1\kern-4.5pt1}

\newcommand{\gam}{\gamma}

\newcommand{\psibar}{\overline{\psi}}

\newcommand{\pslash}{{\not\!\!p}}
\newcommand{\qctd}{QC$_2$D}
\newcommand{\ctcqcd}{\cite{Hands:2006ve}}
\newcommand{\cdse}{\cite{Roberts:2000aa,Nickel:2006vf}}
\newcommand{\cgluon}{\cite{Leinweber:1998uu}}
\newcommand{\cquark}{\cite{Skullerud:2001aw}}
\newcommand{\cgluons}{\cite{Leinweber:1998uu,Cucchieri:2007md,Bogolubsky:2007ud}}

\newcommand{\ckstvz}{\cite{Kogut:2000ek}}

\newcommand{\cprev}{\cite{Skullerud:2008wu}}

\title{Gluons, quarks and deconfinement at high density}

\ShortTitle{Gluons, quarks and deconfinement at high density}

\author{\speaker{Jon-Ivar Skullerud}\\
        NUI Maynooth\\
        E-mail: \email{jonivar@skullerud.name}}

\abstract{We compute gluon and quark propagators in 2-colour QCD at
  large baryon chemical potential $\mu$.  The gluon propagator is
  found to be antiscreened in the superfluid, confined phase and
  screened in the large-$\mu$, deconfined phase.  We present the first
  attempt to determine corresponding electric and magnetic gluon
  masses.  The quark propagator undergoes dramatic modifications in
  the superfluid region as a result of the formation of a superfluid
  gap.  These modifications include the appearance of zero crossings
  in the vector part of the (normal) quark propagator, a large
  suppression of the scalar part, and the emergence of anomalous
  propagation.}

\FullConference{International Workshop on QCD Green's Functions,
		 Confinement and Phenomenology\\ 
		 September 7-11 , 2009\\
		 ECT Trento, Italy}

\begin{document}
\bibliographystyle{JHEP-2}

\section{Introduction}

Determining the phase diagram of QCD at large baryon density and small
temperatures remains one of the outstanding problems of strong
interaction physics.  This problem is of both theoretical and
phenomenological interest: on the theoretical side, an exceptionally
rich phase structure may be present, while the phenomenological
interest is spurred by the possibility that some of these phases may
be present in compact stars, and may have observable consequences.

Direct lattice simulations of QCD at high density and low temperature
are hindered by the sign problem, so alternative approaches are
required. One such approach is to study QCD-like theories which may be
simulated on the lattice, and apply the lessons learnt from these
theories to the case of real QCD.  Foremost among these theories is
QCD with gauge group SU(2) (\qctd).

Medium modifications of quark and gluon propagators is one topic where
\qctd\ may directly inform real QCD calculations.  The gluon
propagator is used as input into the gap equation for the superfluid
gap at high density, but the propagator that is used is usually based
either on (resummed) perturbation theory or on simple generalisations
of the vacuum propagator.
Nontrivial medium modifications or nonperturbative effects may thus
significantly alter the results.  The quark propagator encodes
information about effective quark masses and gap parameters, while
first-principles results for gluon and quark propagators together can
be used to check the assumptions going into dense QCD calculations in
the Dyson--Schwinger equation framework \cdse.


\section{Formulation}

We will be using $N_f=2$ degenerate flavours of Wilson fermion, with a
diquark source $j$ included to lift low-lying eigenvalues and study
diquark condensation without uncontrolled approximations.  The fermion
action can be written
\begin{equation}
S_F = \begin{pmatrix}\psibar_1 & \psi_2^T\end{pmatrix}
\begin{pmatrix} M(\mu) & j\gam_5 \\ -j\gam_5 & M(-\mu)\end{pmatrix}
\begin{pmatrix}\psi_1 \\ \psibar_2^T\end{pmatrix}
\equiv \overline\Psi\mathcal{M}(\mu)\Psi\,,
\end{equation}
where $M(\mu)$ is the usual Wilson fermion matrix with chemical
potential $\mu$.  It satisfies the symmetries %
\begin{align}
KM(\mu)K^{-1} = M^*(\mu)\,,\quad \gam_5M^\dagger(\mu)\gam_5 =
M(-\mu)\,,
\end{align}
with $K=C\gam_5\tau_2$.  The first of these is the Pauli--G\"ursey
symmetry. The inverse of $\mathcal{M}$ is the Gor'kov propagator, %
\begin{equation}
\mathcal{G}(x,y) = \mathcal{M}^{-1} =
\begin{pmatrix}
\bra\psi_1(x)\psibar_1(y)\ket &
\bra\psi_1(x)\psi_1^T(y)\ket\\ \bra\psibar_2^T(x)\psibar_1(y)\ket &
\bra\psibar_2^T(x)\psi_1^T(y)\ket
\end{pmatrix}
=
\begin{pmatrix}S(x,y) & T(x,y) \\ \bar{T}(x,y) & \bar{S}(x,y)
\end{pmatrix} \,.
\end{equation}
The components $S$ and $T$ denote normal and anomalous propagation
respectively.  The Gor'kov propagator has the symmetry properties 
\begin{gather} 
K\mathcal{G}K^{-1} = 
 \begin{pmatrix} S^* & -T^* \\ -\bar{T}^* & \bar{S}^*\end{pmatrix}  \,,\\ 
 \bar{S}(x,y) = -S(y,x)^T\,,\qquad T(x,y)=T(y,x)^T\,,\qquad 
 \bar{T}(x,y) = \bar{T}(y,x)^T\,.  
\end{gather} 
We will also write the inverse
propagator as 
\begin{equation} 
\mathcal{G}^{-1} = \begin{pmatrix} N & \Delta \\ \bar{\Delta} & \bar{N} 
\end{pmatrix}\,, 
\end{equation} 
which has the same symmetry properties as $\mathcal{G}$.  

The normal propagator $S$ can in general
be written in terms of four momentum-space form factors, 
\begin{equation}
S(p) = \not\!\!\vec{p}S_a(\vec{p}^2,p_4) + S_b(\vec{p}^2,p_4) +
\gam_4(p_4-i\mu)S_c(\vec{p},p_4) + i\gam_4\not\!\!\vec{p}S_d(\vec{p}^2,p_4)\,.
\label{formfactors}
\end{equation}
In \qctd\ the Pauli--G\"ursey symmetry ensures that
all form factors are purely real.  The structure of the anomalous
propagator depends on the pattern of diquark condensation.   Assuming
that the condensation occurs in the colour singlet channel with quarks
of unequal flavour, the anomalous propagator can be written as $T(p) =
T'(p)C\Gamma\tau_2$ (and similarly for the anomalous part $\Delta(p)$ of
the inverse propagator), where $\Gamma=\gam_5$ for condensation in the
scalar ($0^+$) channel. 
Spin-1 condensation leads to more complicated
structures, but is energetically disfavoured compared to spin-0
condensation and will not be considered here.   The remaining spin
structure can be written in terms of form factors $T_a, T_b, T_c, T_d$
analogous to \eqref{formfactors}, ie
\begin{equation}\label{anom-formfactors}
T'(p) = \not\!\!\vec{p}T_a(\vec{p}^2,p_4) + T_b(\vec{p}^2,p_4) +
\gam_4(p_4-i\mu)T_c(\vec{p},p_4) + i\gam_4\not\!\!\vec{p}T_d(\vec{p}^2,p_4)\,.
\end{equation}
Similarly, the inverse propagator can be written in terms of form
factors $A, B, C$ and $D$ for the normal part $N$, and $\phi_a,
\phi_b, \phi_c, \phi_d$ for the anomalous part $\Delta'(p)$.  The form
factors $\phi_i$ are the gap functions.

The gluon propagator in presence of a chemical potential in Landau
gauge may be decomposed into an magnetic and electric form factor, 
\begin{equation}
D_{\mu\nu}(\vec{q},q_0) = P^T_{\mu\nu}D_M(\vec{q}^2,q_4^2) +
P^E_{\mu\nu}D_E(\vec{q}^2,q_4^2) \,.
\end{equation}
The projectors $P^T_{\mu\nu}(q), P^E_{\mu\nu}(q)$ are both
4-dimensionally transverse, and are spatially transverse and
longitudinal respectively.  

\section{Results}

We have generated gauge configurations on two lattices: a ``coarse''
lattice with $\beta=1.7, \kappa=0.178, V=8^3\times16$, and a  ``fine''
lattice with $\beta=1.9, \kappa=0.168, V=12^3\times24$.  The lattice
spacings are 0.23fm and 0.18fm respectively, while $m_\pi/m_\rho=0.8$
in both cases.  A range of chemical potentials $\mu$ were used with
diquark source $aj=0.04$, while additional configurations were
generated with $aj=0.02, 0.06$ for selected values of $\mu$.
In addition to this, we have also generated configurations at $\mu=0$
for two ``finer'' lattices, with $\beta=2.0$, $\kappa=0.162$
(``heavy'') and $\kappa=0.163$ (``light''), both with volumes
$V=12^3\times24$. 

\subsection{Gluons}

\begin{figure}[htb]
\begin{center}
\includegraphics*[width=0.8\textwidth]{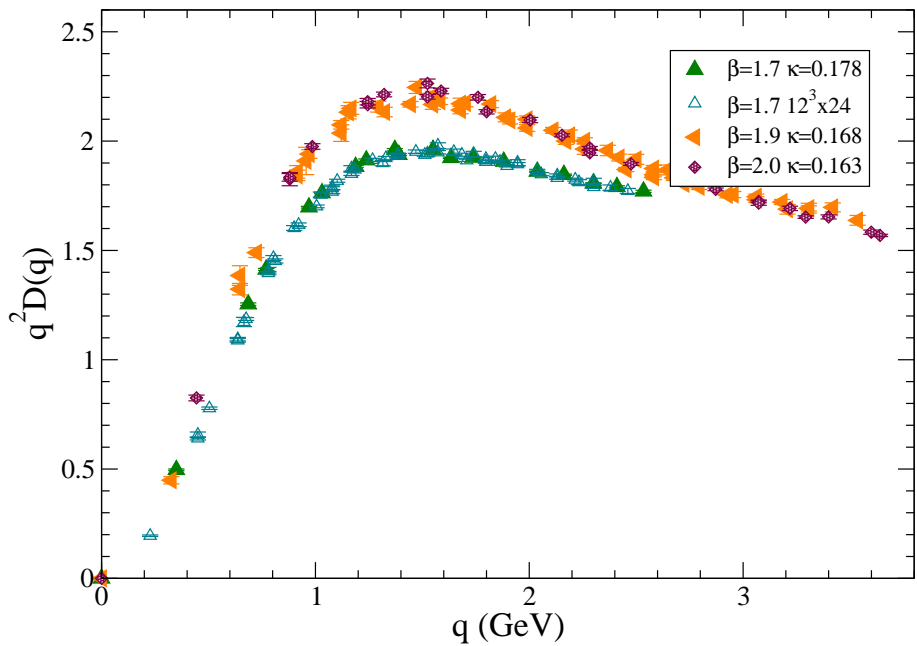}
\end{center}
\caption{The gluon dressing function at zero chemical potential, for
  different lattice spacings and volumes.}\label{fig:gluon-scale}
\end{figure}

Results for the gluon propagator on the coarse lattice have been
presented in \ctcqcd; we will supplement those here with results from
the fine lattice.  On both lattices, an onset transition to a phase
with nonzero baryon density and diquark condensate was found at
$\mu_o\approx m_\pi/2$, while BCS-like scaling of energy density,
baryon density and diquark condensate was found at higher $\mu$.  On
the coarse lattice the crossover to BCS-like scaling was associated
with a nonvanishing Polyakov loop $L$, indicating a coincident
deconfinement transition \ctcqcd.  On the fine lattice, these two
transitions are separate, with the deconfining transition occuring at
considerably larger $\mu$ \cite{Hands:qcd-tnt}.

First of all, we investigate the scaling behaviour of the gluon
propagator in the vacuum ($\mu=0$).  Figure~\ref{fig:gluon-scale}
shows the gluon dressing function $q^2D(q)$ for three of our four
different lattices.  For the coarse lattice parameters, we also have
data for two different volumes.  The data have all been cylinder cut
\cgluon\ to select the points with smallest lattice artefacts.  Since
the lattice spacing for the finer lattice has not yet been
independently determined, the matching procedure described in
\cgluon\ has been used to find the ratio of lattice spacings
$a_f/a_{ff}$ that gives the best match for the gluon propagator on the
fine ($f$) and finer ($ff$) lattices.

We see that finite volume effects are small for the momenta considered
here, but scaling violations (finite lattice spacing effects) are very
large between the coarse and the two finer lattices.  The good scaling
observed between the two finer lattices may be somewhat misleading,
since the matching procedure used in setting the scale for the finer
lattice assumes we are in the scaling r\'egime.  Nonetheless, the good
agreement over a wide range of momenta indicates that lattice
artefacts here are not too large.

\begin{figure}[thb]
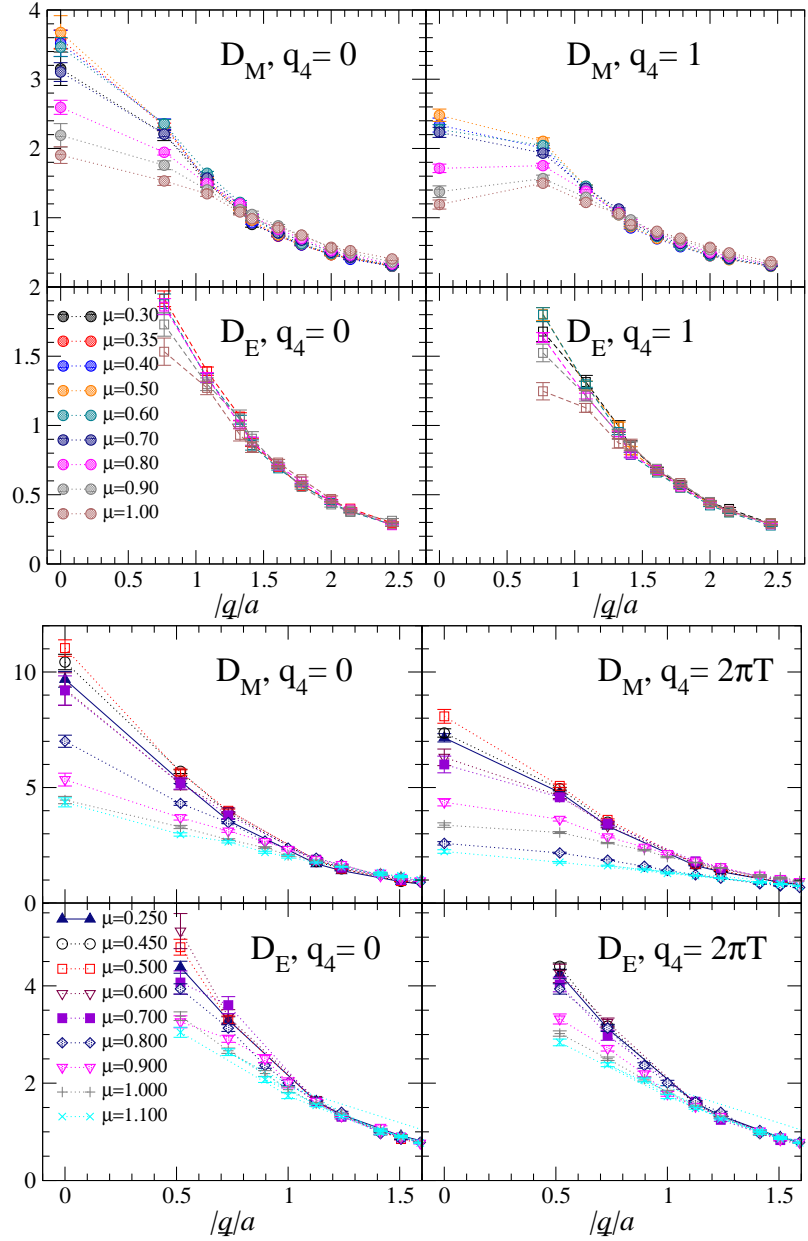

\begin{center}
\includegraphics*[width=0.7\textwidth]{gluonc_k.eps}\\
\includegraphics*[width=0.7\textwidth]{gluonf_k.eps}
\caption{The unrenormalised gluon propagator on the coarse lattice
  (top) and on the fine lattice (bottom), for various chemical
  potentials $a\mu=0.25-1.10$.}
\label{fig:gluon}
\end{center}
\end{figure}
\begin{figure}[thb]
\begin{center}
\includegraphics*[width=0.9\textwidth]{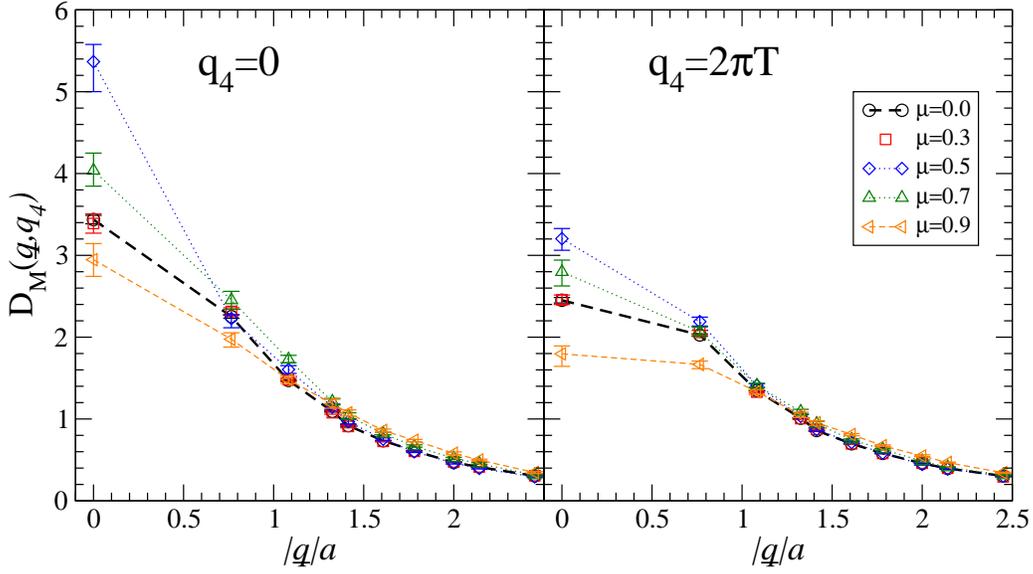}
\caption{Magnetic gluon propagator on the coarse lattice, extrapolated
  to zero diquark source $j$.  The left-hand plot shows the lowest
  Matsubara mode ($q_4=0$), while the right-hand plot shows the first
  nonzero Matsubara mode.}
\label{fig:DMc_j0}
\end{center}
\end{figure}
Figure~\ref{fig:gluon} shows the two lowest Matsubara modes of the
unrenormalised gluon propagator as a function of spatial momentum
$|\vec{q}|$ for a range of chemical potentials, on both lattices.  In
all cases, the propagator at the lowest chemical potential $\mu$ shown
is consistent with the vacuum propagator.  On the coarse lattice both
magnetic and electric propagator are strongly screened at large $\mu$,
while they are enhanced at low momentum in the intermediate-density
region.  The static ($q_0=0$) magnetic gluon propagator turns out to
have a surprisingly strong dependence on the diquark source, which
counteracts the infrared suppression at large $\mu$ as $j\to0$, but
does not remove it completely.  This is demonstrated in
fig.~\ref{fig:DMc_j0}, which shows the magnetic gluon propagator for
the two lowest Matsubara frequencies, extrapolated to zero diquark
source.  We clearly see a strong infrared enhancement at intermediate
$\mu$, but at $a\mu=0.9 (\mu=0.78$GeV) both the static and non-static
modes are suppressed in the infrared.

The same qualitative picture can be seen on the fine lattice, but in
this case the infrared suppression sets in at much larger $\mu$
(around $a\mu=0.8$ or $\mu=0.9$GeV).  This is consistent with the
hypothesis that the screening effect is linked with the deconfinement
transition, ie that it is a result of the gluons being screened by
coloured quark degrees of freedom.

It is worth pointing out that the enhancement resp. screening noted
here is in comparison to the vacuum gluon propagator, which is known
to be infrared suppressed due to nonperturbative effects (as discussed
at length in other contributions to this conference).  It seems
reasonable to assume that although the static magnetic gluon is
unscreened to all orders in perturbation theory, nonperturbative
effects may be responsible for the additional screening observed here
in the deconfined phase.

\begin{figure}[hbt]
\begin{center}
\includegraphics*[width=0.8\textwidth]{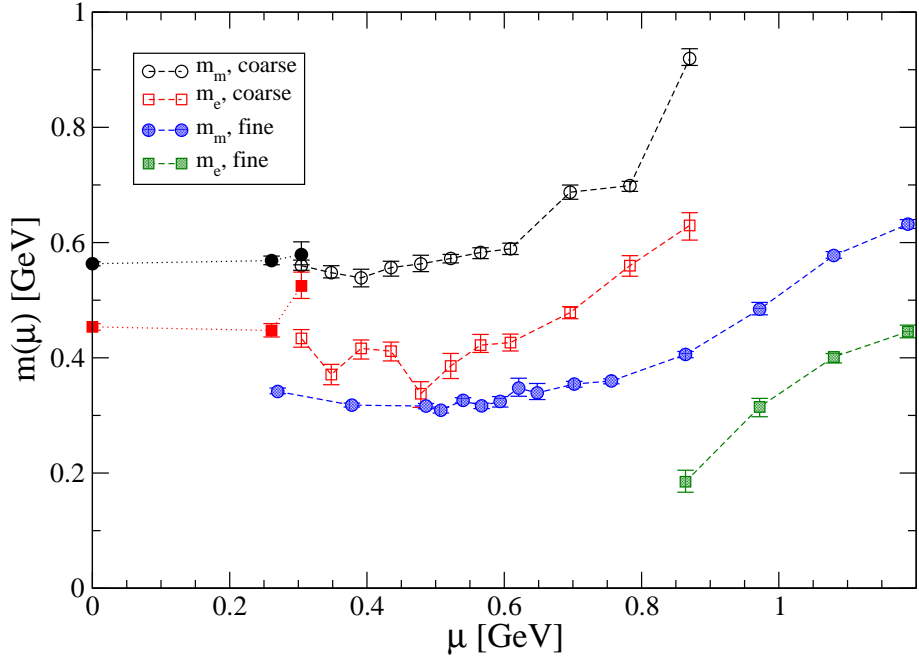}
\end{center}
\caption{The electric and magnetic gluon mass as a function of
  chemical potential $\mu$, determined from a fit to a simple massive
  propagator on each lattice.  For the coarse lattice, the filled
  symbols denotes fits to data with zero diquark source $j$, while the
open symbols are from $ja=0.04$.  For the fine lattice all data are
for $ja=0.04$.  It was not possible to get any fit for the
electric gluon on the fine lattice for $a\mu<0.7$.}
\label{fig:gluonmass}
\end{figure}

We have attempted to fit the gluon propagator to a simple massive
form,
\begin{equation}
D_{E,M}(\vec{q},q_4;\mu) = \frac{Z_{e,m}}{\vec{q}^2+q_4^2+m^2_{e,m}(\mu)}\,.
\end{equation}
The resulting electric and magnetic gluon masses $m_{e,m}$ are shown
as functions of $\mu$ in figure~\ref{fig:gluonmass}.  It is worth
noting that the quality of these fits is quite poor.  This is
expected, as it is known that at $\mu=0$ it is not possible to
describe the gluon propagator by a simple, momentum-independent mass,
while at large $\mu$ one should reproduce the results of
hard-dense-loop (HDL) resummed perturbation theory, which also has a more
complicated functional form.  A form which interpolates between HDL
and available results for $\mu=0$ \cgluons\ is likely to yield 
better results.  A further technical complication is that we have
defined $D_E$ only at nonzero spatial momenta, while the fits to $D_M$
include the $\vec{q}=0$ point.  This is the reason for the discrepancy
between $m_e$ and $m_m$ at $\mu=0$, where they should be equal.  This
also tends to yield lower values for $m_e$ throughout.

With these provisos, we can see that both the electric and magnetic
gluon masses remain roughly constant for small and intermediate $\mu$,
before increasing at large $\mu$, corresponding roughly to the
deconfined phase.  We see, however, that there is a large difference
between the mass values from the two lattices, indicating that scaling
violations are still very large at these lattice spacings.

\subsection{Quarks}

\begin{figure}[hbt]
\begin{center}
\includegraphics*[width=0.95\textwidth]{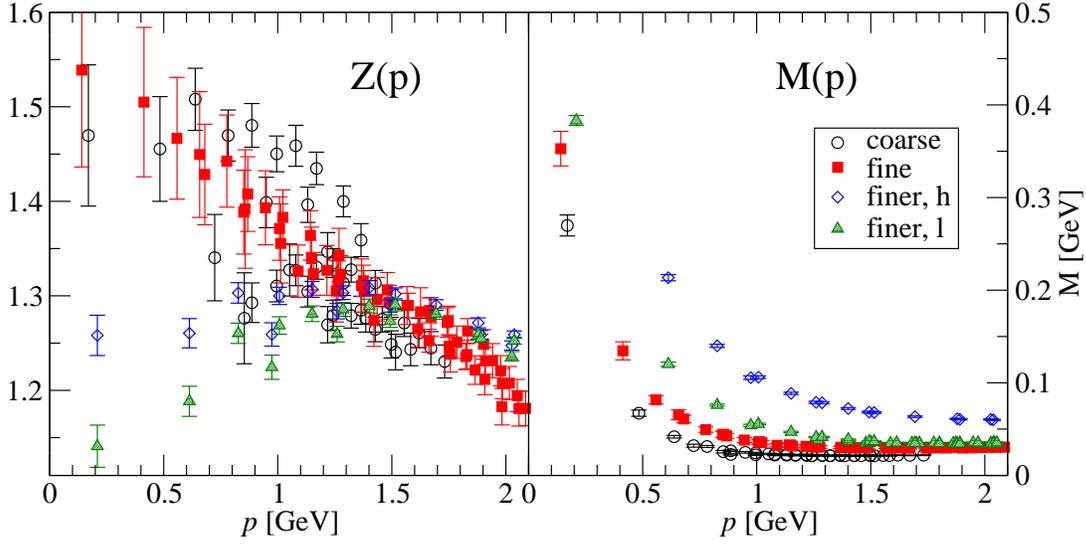}
\end{center}
\caption{The quark propagator renormalisation function (left) and mass
  function (right) at zero chemical potential, for different lattice
  spacings.}
\label{fig:vacuum-quark}
\end{figure}

In the vacuum, there are only two independent tensor components of the
quark propagator, which is conventionally written as
\begin{equation}\label{vacuum-quark}
S(p) = \frac{Z(p)}{i\pslash +M(p)}\,,
\end{equation}
where $M$ is the mass function and $Z$ the renormalisation function.
These are shown in fig.~\ref{fig:vacuum-quark}, for the different
lattice spacings and quark masses available.  Both $Z(p)$ and $M(p)$
have been multiplicatively tree-level corrected \cquark; however, since the
critical quark mass is not yet known, the tree-level correction of
$M(p)$ is not yet properly carried out. 

We immediately see that there are large scaling violations in both
form factors, and large violations of rotational symmetry in $Z(p)$.
In particular, we note that $Z(p)$ increases in the infrared for the
coarser lattices, whereas it is usually found to be infrared
suppressed.  We see that this suppression appears to be recovered as
we move towards the continuum limit.  A careful continuum
extrapolation will be needed to obtain quantitative results.

At nonzero chemical potential, we find that the form factors $S_a,
S_b$ and $S_c$ (spatial-vector, scalar and temporal-vector) of the
normal quark propagator and the form factors $T_b$ and $T_d$ (scalar
and tensor) of the anomalous propagator are nonzero, while the
remaining components are zero.  Results for the coarse lattice were
shown in \cprev; here we will show results for the fine lattice only.

\begin{figure}[hbt]
\begin{center}
\includegraphics*[width=0.9\textwidth]{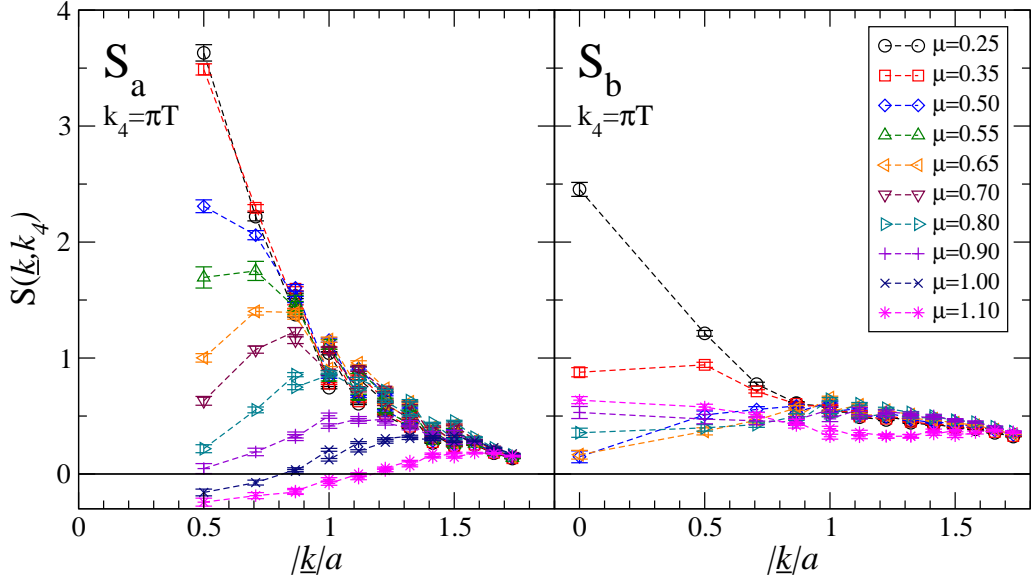}
\end{center}
\caption{The lowest Matsubara frequency of the spatial-vector (left)
  and scalar (right) part of the normal quark propagator, on the fine
  lattice, for different chemical potentials $\mu$.}
\label{fig:normal}
\end{figure}
\begin{figure}[hbt]
\begin{center}
\includegraphics*[width=0.9\textwidth]{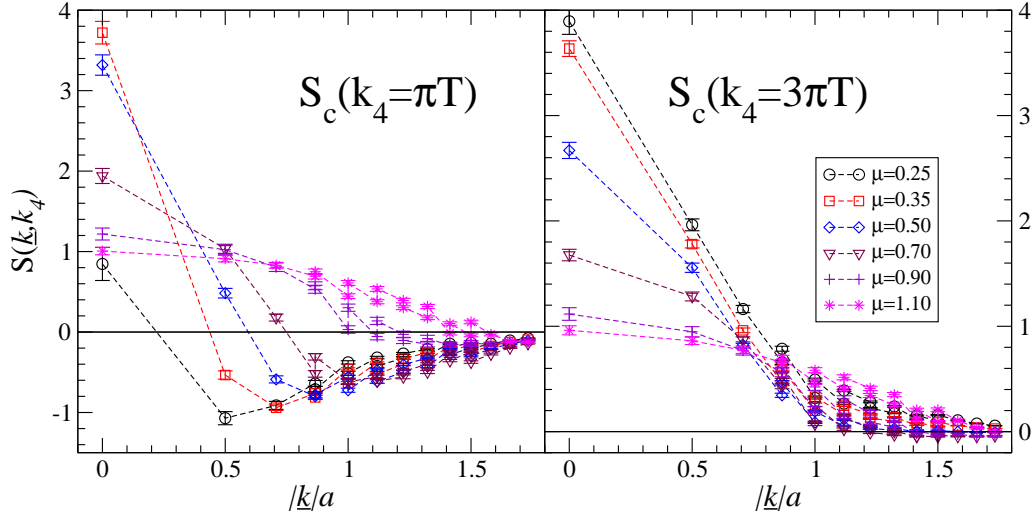}
\end{center}
\caption{The temporal-vector part of the
  normal quark propagator, on the fine lattice, for different chemical
  potentials $\mu$.}
\label{fig:normal-c}
\end{figure}
Figure~\ref{fig:normal} shows the spatial-vector part $S_a$ and scalar
part $S_b$ of the normal quark propagator for a range of chemical
potentials $a\mu=0.25-1.1$. These both exhibit dramatic
medium modifications.  The scalar propagator $S_b$ is strongly
suppressed in the superfluid phase, suggesting a drastic reduction in
the in-medium effective quark mass.  This is linked to the appearence
of the diquark condensate: the chiral condensate rotates into the
diquark condensate in the superfluid phase \ckstvz.  We would
therefore expect to find the missing strength in the anomalous
propagator.  The change in behaviour is sudden and takes place around
$\mu_o=m_\pi/2$, while for larger $\mu$ there is little change.

The spatial-vector propagator $S_a$ is also infrared suppressed at
large $\mu$, but this suppression happens gradually as a function of
$\mu$, and sets in only above $\mu_o$.  At the largest densities we
see that $S_a(\vec{k},k_4=\pi T)$ becomes negative for small spatial
momentum $|\vec{k}|$.

The two lowest Matsubara modes of the temporal-vector propagator $S_c$
are shown in \ref{fig:normal-c}. We see that the lowest Matsubara mode
($k_4=\pi T$) becomes negative at intermediate momenta, approaching
zero from below at high momenta.  This is a dramatic change compared
to the vacuum propagator, which stays positive at all momenta, and
indicates the formation of a superfluid gap.  The location of the zero
crossing in the $k_4\to0$ limit corresponds to the Fermi
momentum $k_F$.  In accordance with this, the zero crossing moves to
larger $|\vec{k}|$ as $\mu$ increases.  On closer inspection, we find
that the second Matsubara mode ($k_4=3\pi T$) also becomes negative
for large $\mu$ (at large spatial momentum).  It would therefore in
principle be possible to extrapolate this zero crossing to $k_4=0$ and
thus find $k_F$ as a function of $\mu$.

\begin{figure}[hbt]
\includegraphics*[width=0.9\textwidth]{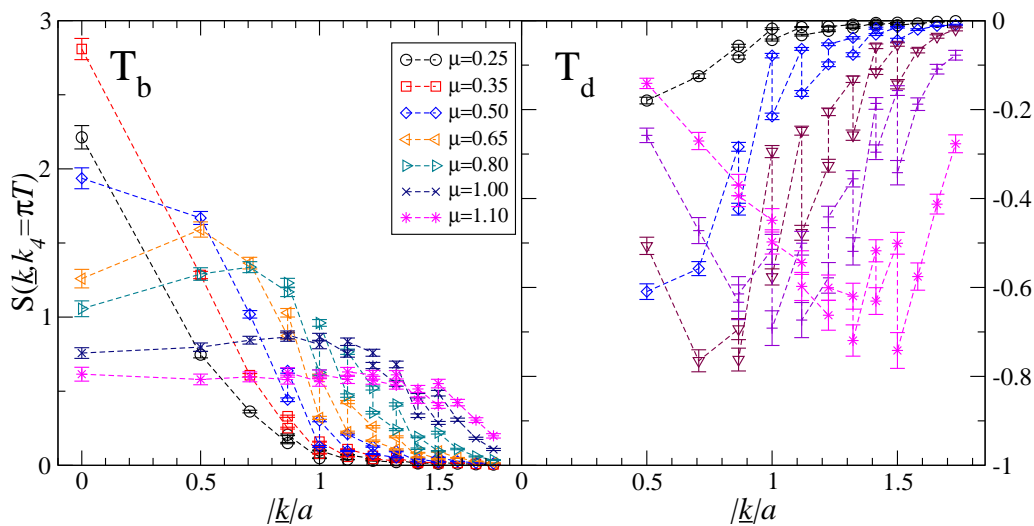}
\caption{The scalar (left) and tensor (right) part of the anomalous
  quark propagator, on the fine lattice.}
\label{fig:anomalous}
\end{figure}
Figure~\ref{fig:anomalous} shows the nonzero components of the
anomalous Gor'kov propagator.  The dominant part is, as expected, the
scalar part $T_b$, but a clear signal is also found for the tensor
part $T_d$, in accordance with what was found on the coarse lattice
\cprev.  We find that the lattice artefacts in the scalar part is
substantially reduced compared to the coarse lattice, while the tensor
part is still subject to very large violations of rotational
symmetry.  It may therefore be open to question whether this component
will survive the continuum limit.

The scalar anomalous propagator shows a clear change in behaviour as
one goes from small to large chemical potential.  Firstly, we note
that it increases between $a\mu=0.25$ and 0.35.  The former point is
below the superfluid transition, but anomalous propagation is present
due to the explicit diquark source.  We expect that $T_b$ (and $T_d$)
will vanish in the $j\to0$ limit for $\mu<\mu_o$.  As $\mu$ increases
above $\mu_o$, $T_b$ develops a plateau at low momentum, which extends
to larger $|\vec{k}|$ with increasing $\mu$.  At large $\mu$, $T_b$
thus becomes approximately constant, suggesting that anomalous
propagation may be described by a momentum-independent diquark gap
$\Delta$.

\section{Discussion and outlook}

We have found substantial modifications of both gluon and quark
propagators in the dense medium.  In the superfluid, confined phase,
the electric and magnetic gluon propagators are both enhanced in the
infrared compared to the vacuum.  In the deconfined phase, they are
both screened (infrared suppressed).  This screening is evident even
in the static magnetic gluon, which is unscreened to all orders in
perturbation theory.  If these results carry over to SU(3) they would
invalidate the use of an unscreened static-magnetic gluon propagator
in the gap equation at large $\mu$.  A careful analysis at different
volumes and lattice spacings is however necessary to draw quantitative
conclusions.

The dramatic modifications seen in the quark propagator are directly
related to the appearance of a diquark gap.  Our next step will be to
compute the form factors, including the diquark gap and mass
function, by inverting the quark propagator.  Further quantitative
studies will include determining the Fermi momentum $p_F$ by
extrapolating the zero crossing in the temporal-vector propagator
$S_c$ to $k_4=0$, and determining the size of Cooper pairs from the
anomalous propagator, to study the BEC--BCS crossover in more detail.

\section*{Acknowledgments}

I wish to thank the organisers for a very stimulating workshop.  I
also wish to thank Simon Hands for his collaboration in this research,
and Dominik Nickel for very fruitful discussions.

\bibliography{gluon,density,qcd}

\end{document}